\documentclass[twocolumn,showpacs,preprintnumbers,amsmath,amssymb]{revtex4}

\usepackage{epsfig}
\usepackage{psfig}
\usepackage{subfigure}
\usepackage{graphicx}
\usepackage{dcolumn}
\usepackage{bm}
\usepackage{amsmath}
\usepackage{amssymb}

\newcommand{\pp}{{\it \={p}p }}
\newcommand{\cc}{{\it \={c}c }}

\newcommand{\ee}{\mbox{$e^+ e^-~$}}
\newcommand{\jpsi}{\mbox{$J/\psi~$}}
\newcommand{\psip}{\mbox{$\psi^{\prime}~$}}
\newcommand{\gamgam}{\mbox{$\gamma \gamma~$}}

\begin{document}


\title{Measurement of the branching ratios $\psi^\prime \to e^+ e^-$, 
$\psi^\prime \to J/\psi \pi \pi$ and $\psi^\prime \to J/\psi \eta$.}

\author{M.~Andreotti$^{2}$, S.~Bagnasco$^{3,7}$, W.~Baldini$^{2}$, 
D.~Bettoni$^{2}$, G.~Borreani$^{7}$, A.~Buzzo$^{3}$, R.~Calabrese$^{2}$, 
R.~Cester$^{7}$, G.~Cibinetto$^{2}$, P.~Dalpiaz$^{2}$, G.~Garzoglio$^{1}$, 
K.~E.~Gollwitzer$^{1}$, M.~Graham$^{5}$, M.~Hu$^{1}$, D.~Joffe$^{6}$, 
J.~Kasper$^{6}$, G.~Lasio$^{4}$, M.~Lo~Vetere$^{3}$, E.~Luppi$^{2}$, 
M.~Macr\`\i$^{3}$, M.~Mandelkern$^{4}$, F.~Marchetto$^{7}$, M.~Marinelli$^{3}$, 
E.~Menichetti$^{7}$, Z.~Metreveli$^{6}$, R.~Mussa$^{7}$, M.~Negrini$^{2}$, 
M.~M.~Obertino$^{7}$, M.~Pallavicini$^{3}$, N.~Pastrone$^{7}$, C.~Patrignani$^{3}$, 
T.~K.~Pedlar$^{6}$, S.~Pordes$^{1}$, E.~Robutti$^{3}$, W.~Roethel$^{6,4}$, 
J.~L.~Rosen$^{6}$, P.~Rumerio$^{6}$, R.~Rusack$^{5}$, A.~Santroni$^{3}$, 
J.~Schultz$^{4}$, S.~H.~Seo$^{5}$, K.~K.~Seth$^{6}$, G.~Stancari$^{1}$, 
M.~Stancari$^{4}$, A.~Tomaradze$^{6}$, I.~Uman$^{6}$, T.~Vidnovic~III$^{5}$, 
S.~Werkema$^{1}$ and P.~Zweber$^{6}$ \\
(FNAL experiment E835)}

\affiliation{$^{1}$Fermi National Accelerator Laboratory, Batavia, Illinois 60510 \\
$^{2}$Istituto Nazionale di Fisica Nucleare and University of Ferrara, 44100 Ferrara, Italy \\
$^{3}$Istituto Nazionale di Fisica Nucleare and University of Genova, 16146 Genova, Italy \\
$^{4}$University of California at Irvine, Irvine, California 92697 \\
$^{5}$University of Minnesota, Minneapolis, Minnesota 55455 \\
$^{6}$Northwestern University, Evanston, Illinois 60208 \\
$^{7}$Istituto Nazionale di Fisica Nucleare and University of Torino, 10125 Torino, Italy }

%

\date{\today}

\begin{abstract}
We have measured several branching ratios for \psip decay using the 
data collected by FNAL E835 experiment during year 2000, obtaining 
${\cal B}(\psi^\prime \to e^+ e^-) = 0.0068\pm0.0001\pm0.0004$,
${\cal B}(\psi^\prime \to J/\psi \pi^+ \pi^-) = 0.292\pm0.005\pm0.018$, 
${\cal B}(\psi^\prime \to J/\psi \pi^0 \pi^0) = 0.167\pm0.005\pm0.014$ and 
${\cal B}(\psi^\prime \to J/\psi \eta) = 0.028\pm0.002\pm0.002$. 
We also present a measurement of the dipion mass distribution in the 
decays $\psi^\prime \to J/\psi \pi \pi$. 
\end{abstract}

\pacs{13.75.Cs, 14.40.Gx}
\maketitle

\section{Introduction} \label{introduction_sec}
Branching ratios of \psip decay have been measured by many experiments 
in \ee collisions. The branching ratios to $e^+ e^-$ and to 
\jpsi inclusive states have been measured also by the Fermilab 
experiments E760 \cite{Armstrong:pg} and E835 \cite{Ambrogiani:2000vj}, 
which studied charmonium spectroscopy in \pp annihilation. 

The strong \psip decays to \jpsi proceed mainly through the 
emission of soft gluons by the \cc pair, with their subsequent hadronization 
\cite{Goldberg:1975bz}. 
The low gluon momentum makes this process nonperturbative. 
However some features of the hadronic process 
$\psi^\prime \to J/\psi X$ can be predicted using conservation laws. 
A prediction based on isospin conservation is that 
${\cal B}(\psi^\prime \to J/\psi \pi^0 \pi^0) / {\cal B}(\psi^\prime \to J/\psi \pi^+ \pi^-) = \frac{1}{2}$. 
While previous measurements have always yielded slightly larger values, 
further checks of this ratio are needed. 
The multipole expansion of the gluon field \cite{Gottfried:1977gp} has 
been used to predict a value in the range 0.10 - 0.14 for 
$\Gamma(\psi^\prime \to J/\psi \eta) / \Gamma(\psi^\prime \to J/\psi \pi^+ \pi^-)$ 
\cite{Voloshin:1980zf,Novikov:fa} and the $\pi\pi$ 
invariant mass distribution \cite{Pham:1975zq,Yan:1980uh}.


\section{Experimental method} \label{experiment_sec}
E835 studies charmonium spectroscopy in 
\pp annihilation. All \cc states can be directly formed with this technique. 
Since the cross section for charmonium formation is at best $\sim 5$ 
orders of magnitude smaller than the total \pp cross section, in order 
to improve the signal to noise ratio the charmonium signal is extracted 
from the hadronic background by detecting electromagnetic final states. 

\psip decays are studied here by selecting events with a high invariant mass 
\ee pair in the final state, coming from the reactions:
\begin{eqnarray}
\psi^\prime &\to& e^+ e^- , \nonumber \\
\psi^\prime &\to& J/\psi \pi^+ \pi^- \to e^+ e^- \pi^+ \pi^- , \nonumber \\
\psi^\prime &\to& J/\psi \pi^0 \pi^0 \to e^+ e^- 4 \gamma , \nonumber \\
\psi^\prime &\to& J/\psi \eta \to e^+ e^- 2 \gamma . \nonumber 
\end{eqnarray}

The number $N_A$ of events observed for \psip decay to a given final state 
$A$ is given by: 
\begin{equation}
N_A = {\cal L} \cdot \left( \sigma_{bkg} + \epsilon_A \cdot \int G(E) \cdot \sigma_{BW}(E) \cdot {\cal B}(\psi^\prime \to A) dE \right) ,
\label{numevents_fmla}
\end{equation}
where ${\cal L}$ is the integrated luminosity, $\sigma_{bkg}$ the background 
cross section, $\epsilon_A$ the overall detection efficiency for the channel, 
$\sigma_{BW}$ the Breit-Wigner cross 
section for $\overline{p}p \to \psi^\prime$, $G(E)$ the center of mass 
energy distribution and ${\cal B}(\psi^\prime \to A)$ the 
branching ratio for the decay $\psi^\prime \to A$. 

Measuring the ratio of branching ratios for two channels $A$ and $B$ on the 
same data sample, several factors in (\ref{numevents_fmla}) cancel, 
leading to:
\begin{equation}
\frac{{\cal B}(\psi^\prime \to A)}{{\cal B}(\psi^\prime \to B)} = 
    \frac{\epsilon_B}{\epsilon_A} \cdot \frac{N_A - N_{bkg,A}}{N_B - N_{bkg,B}} ,
\label{fractevents_fmla}
\end{equation}
where $N - N_{bkg}$ is the number of events after the subtraction of 
the background contribution. 
The reference channel has been chosen to be the $J/\psi$ inclusive decay 
because it is the one for which the ratio (\ref{fractevents_fmla}) presents 
the lowest systematics, as will be shown in Section \ref{effic_sec}. 


\section{Experimental apparatus} \label{detector_sec}
E835 \cite{det-paper} is a fixed target experiment in which the $\overline{p}$ 
beam circulating in the Fermilab Antiproton Accumulator (AA) crosses an internal 
hydrogen gas jet target. 
During a typical data taking period the $\overline{p}$ are accumulated 
until a current of $\sim 70$ mA is reached.
Then accumulation stops and the $\overline{p}$ are stochastically cooled 
and decelerated to the energy of the resonance to be studied. 
At this point the H$_2$ target is turned on and the data taking proceeds 
on the resultant ``stack'' until the $\overline{p}$ beam current is around 
10 mA. In this current range it is possible to take data at constant luminosity 
by regulating the jet target density. 

After deceleration, the mean center of mass energy $E_{cm}$ of the \pp 
system is known with a precision better than $\sim 100$ keV, where the 
$E_{cm}$ distribution is gaussian with a $\sigma$ typically ranging from 
200 to 400 keV. 
Further details about the AA operation during E835 data taking can be found 
in \cite{McGinnis:gt}. 

The E835 detector is a nonmagnetic 
spectrometer with cylindrical symmetry around the beam axis. 
The inner part of the detector is the charged tracking system; 
for the year 2000 data taking it 
was composed of three cylindrical hodoscopes, two straw chambers for the 
measurement of the azimuthal angle $\phi$ (around the beam axis), two 
scintillating fiber detectors for the measurement of the polar angle 
$\theta$ (with respect to the beam axis) and an additional hodoscope 
in the forward direction used as a charged veto. 
The three cylindrical hodoscopes are segmented in $\phi$ in 8 
(for the inner), 24 (for the intermediate) and 32 (for the outer) 
modules respectively. The hodoscopes provide $dE/dx$ information for the 
charged tracks and they are used to form the charged hardware trigger. 

A 16 cell \v{C}erenkov counter with 8 azimuthal and two polar angle 
sections cover the full azimuth and the polar angle region
$15^\circ \leq \theta \leq 65^\circ$ and allows the separation of high energy 
$e^\pm$, mostly produced in \jpsi and \psip decay, from the other 
charged particles, which is used in the first level trigger. 

Two electromagnetic calorimeters, the Central Calorimeter (CCAL) and 
the Forward Calorimeter (FCAL) cover the region 
$11^\circ \leq \theta \leq 70^\circ$ and $3^\circ \leq \theta \leq 11^\circ$ 
respectively.
The CCAL energy resolution is $\sigma_E/E = 1.4\% + 6\% / \sqrt{E(GeV)}$ 
and its angular resolutions are $\sigma_\phi \sim 11$ mrad and 
$\sigma_\theta \sim 6$ mrad. 
FCAL is not used in the following analysis. 

The luminosity is measured with $\sim 2\%$ accuracy by means of solid state 
detectors located below the interaction region at $\theta \simeq 90^\circ$ to 
the $\overline{p}$ beam.

\section{Event selection} \label{evtsel_sec}
In year 2000, E835 collected a total integrated luminosity of 
${\cal L} = 113 $ pb$^{-1}$ of data. 
The data sample used in this analysis consists of 
14.4 pb$^{-1}$ collected in the \psip resonance region, with 
12.5 pb$^{-1}$ on the resonance and 1.9 pb$^{-1}$ 
in two samples at energies above (3704.9 MeV) and below 
(3666.1 MeV) the \psip, used for background measurements. 

Interesting events are characterized by two nearly back-to-back 
high energy $e^\pm$. The first level hardware trigger is the 
logical OR of one main trigger condition and two control 
triggers \cite{Baldini:wg}.
The main hardware trigger requires two ``electron tracks'' 
as defined by signals in the \v{C}erenkov and corresponding hodoscope elements
in coincidence with two back to back energy deposits in the calorimeter.
The efficiency of the main trigger was measured on data taken with a 
dedicated trigger, and continuously monitored by the two control triggers, 
the first with no \v{C}erenkov requirement and the second with no 
calorimeter requirement. 

As a preliminary selection, all events with \ee candidates with invariant 
mass $m_{ee} < 2.6$ GeV are rejected. 

\begin{figure}
\centering
\epsfig{figure=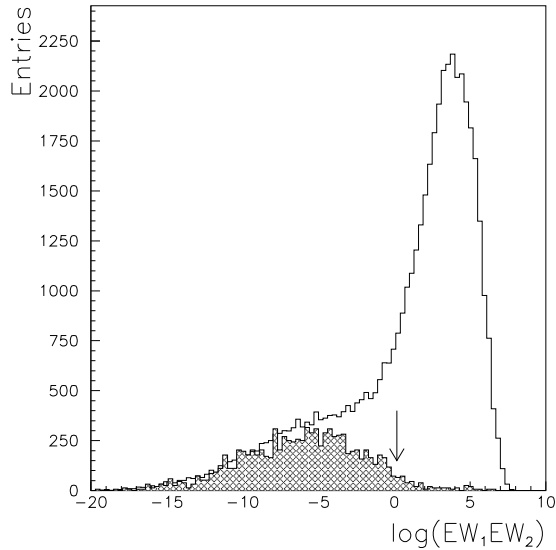,width=.45\textwidth}
\caption{$\log_{10} (EW_1 \cdot EW_2)$ distribution for events with $m_{ee} > 2.6$ GeV 
in the \psip energy region (open) and on the background with normalized 
luminosity (cross hatched). The cut 
$\log_{10} (EW_1 \cdot EW_2) > \log_{10} (1.5)$ is indicated by the arrow.}
\label{ew_bkg_fig}
\end{figure}

\begin{figure}
\centering
\epsfig{figure=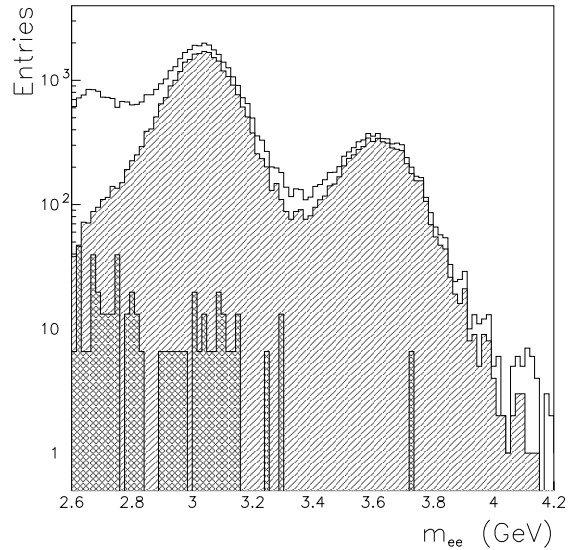,width=.45\textwidth}
\caption{$m_{ee}$ distribution for all the events with candidate \ee pair with 
$m_{ee} > 2.6$ GeV (open). The events selected with the $EW_1 \cdot EW_2 > 1.5$ 
cut in the \psip energy region (hatched) and on the background scaled with the 
luminosity (cross hatched), are also shown.}
\label{gxmtotal_fig}
\end{figure}

A maximum likelihood method called ``Electron Weight'' (EW) 
has been developed for the rejection of backgrounds that mimic $e^\pm$ 
tracks in the detector (mainly $\gamma$ conversion and $\pi^0$ Dalitz decay), 
comparing the signal and background probability for the candidate \ee pair 
on the basis of pulse heights in hodoscopes and \v{C}erenkov counters and 
the CCAL shower shape. Figure \ref{ew_bkg_fig} shows the logarithm of 
the EW product for the two $e^\pm$ candidates for data in 
the \psip energy region and for the background.
We choose to cut at $EW_1 \cdot EW_2 > 1.5$.
In this way a clean sample of \ee due to charmonium decay is selected. 
Figure \ref{gxmtotal_fig} shows the invariant mass distributions of the 
\ee candidates before and after the EW cut. 
After the cut, the two $m_{ee}$ peaks due to the direct decay 
$\psi^\prime \to e^+ e^-$ and the cascade 
$\psi^\prime \to J/\psi X \to e^+ e^- X$ are clearly visible. 

All the events containing an additional shower in the CCAL within 
100 mrad of the $e^\pm$ tracks are rejected. This cut yields homogeneous 
values of the efficiency $\epsilon_{EW}$ for all the final states and 
reduces the systematic error in the final branching ratio measurements. 

$\psi^\prime \to e^+ e^-$ and $\psi^\prime \to J/\psi X$ are first 
selected using kinematical fits requiring ${\rm prob}(\chi^2) > 10^{-4}$; 
the $\psi^\prime \to J/\psi X$ are then tagged as $J/\psi \pi^+ \pi^-$, 
$J/\psi \pi^0 \pi^0$ or $J/\psi \eta$ with kinematical fit selections 
using the information on the additional tracks in the detector. 

For $J/\psi \pi^+ \pi^-$ the $\pi^\pm$ tracks are reconstructed 
within the charged tracking system acceptance 
($15^\circ \leq \theta_{\pi^\pm} \leq 55^\circ$) 
combining the $\phi$ measurements obtained with the straw chambers 
($\phi$-lines) and the $\theta$ measurements obtained with 
the scintillating fiber detectors ($\theta$-lines).
For each event containing at least two $\phi$- and two $\theta$-lines 
in addition to $e^+ e^-$ tracks, 
all the possible pairs of charged $\theta$-$\phi$ lines combinations 
are tested with the kinematical fit as $\pi^+ \pi^-$ candidates. 
The combination with the highest $\chi^2$ probability is chosen. 
The $\pi^\pm$ energy cannot be measured in the E835 detector so 
a 3C kinematical fit is applied for 
the selection of $J/\psi \pi^+ \pi^-$ events. 

The $\pi^\pm$ have relatively low momentum in the laboratory and 
the multiple scattering effect dominates the uncertainty in the 
measurement of their directions. 
The associated error has been determined using a full detector simulation 
based on the GEANT \cite{geant} package.

Photons are detected as energy deposits in the CCAL 
fiducial region $12^\circ \leq \theta_\gamma \leq 68^\circ$. 
$\pi^0$ and $\eta$ candidates are sought in their \gamgam decay, 
looking for photon pairs of invariant mass between 50 MeV and 350 MeV or 
300 MeV and 800 MeV respectively. 

The previous final state categories are mutually exclusive. 
If an event passes more than one of the previous selections, the event 
is assigned to the final state with the highest ${\rm prob}(\chi^2)$. 
The number of events selected is reported in Table \ref{nevents_tab}. 
We find that of the 31193 candidate events, 30492 (98\%) are selected 
as \ee or $J/\psi X$, in good agreement with the Monte Carlo results for 
the efficiencies of the two selections, as described in Section \ref{effic_sec}. 

\begin{table*}
\caption{Number of events observed in each decay channel in E835 year 2000 
\psip data, divided by stack. Stacks 54a and 54c are collected far from 
the resonance and are used for the background measurements. 
The charged tracking system was off for part of the stack 54a.
During this period the $J/\psi \pi^+ \pi^-$ could not be detected so 
the effective total integrated luminosity on the background (stacks 54a 
and 54c) for this channel is 1.7 pb$^{-1}$.} 
\label{nevents_tab}
\begin{ruledtabular}
\begin{tabular}{cccccccc}
Stack & ${\cal L}$ (pb$^{-1}$) & Candidates & $e^+ e^-$ & $J/\psi X$ & $J/\psi \pi^+ \pi^-$ & $J/\psi \pi^0 \pi^0$ & $J/\psi \eta$  \\
\hline
 1  & 0.748 &   831 &  155 &  658 & 114 &  31 & 12  \\
 2  & 1.008 &  3595 &  636 & 2893 & 511 & 187 & 41  \\
14  & 0.992 &  2248 &  388 & 1807 & 323 & 132 & 27  \\
29  & 0.992 &  2083 &  313 & 1706 & 273 & 126 & 17  \\
30  & 0.396 &   951 &  158 &  769 & 135 &  45 &  9  \\
49  & 2.566 &  5890 &  931 & 4840 & 781 & 322 & 67  \\
50  & 1.275 &  3633 &  597 & 2966 & 385 & 205 & 43  \\
51  & 2.103 &  5016 &  812 & 4080 & 660 & 275 & 65  \\
54b & 2.401 &  6946 & 1114 & 5669 & 971 & 373 & 82  \\
\hline
54a & 1.153 &	 35 &	 1 &   29 &	1 &   1 &  0  \\
54c & 0.780 &	 23 &	 0 &   19 &	1 &   1 &  0  \\
\end{tabular}
\end{ruledtabular}
\end{table*}

\section{Background subtraction} \label{backgnd_sec}
Two sources of background must be subtracted from 
the observed numbers of events: a non resonant 
background contribution and an internal background due to 
final state misidentification. 

The non resonant (external) background is due to events not coming from 
\psip decay. 
It is measured on data taken in the energy regions 
far from the resonance (stacks 54a and 54c shown in Table \ref{nevents_tab}) 
by applying the same analysis as for the \psip peak data; 
its contribution is compared to the resonance signal in Figure \ref{gxmtotal_fig}. 
The number of background events from this source to be subtracted is obtained 
scaling the number of events observed in each sample with the integrated 
luminosity. 

The misidentification background is due to resonant events which are 
not correctly classified. 
It is evaluated using a Monte Carlo simulation: a sample of events with 
full detector simulation is generated for all the contaminating channels and 
analyzed with the same analysis used for the data. 
The probability of event misidentification is shown in Table \ref{misidfrac_tab}. 
The number of internal background events to be subtracted for each 
exclusive contaminating channel ($N_{int}(A)$, where $A$ indicates the 
contaminating final state) is obtained from the number of observed $J/\psi X$ 
events, scaled with the branching ratio decay mode obtained from 
the PDG \cite{Hagiwara:pw}, and multiplied by the misidentification probability 
$P_{misid}$ (Table \ref{misidfrac_tab}):
\begin{equation}
N_{int}(A) = N(J/\psi X) \cdot 
  \left[ \frac{{\cal B}(\psi^\prime \to A)}{{\cal B}(\psi^\prime \to J/\psi X)} \right]_{PDG} \cdot
  P_{misid}
\end{equation}

\begin{table*}
\caption{Misidentification probabilities. Contaminations smaller than $10^{-3}$ 
have been neglected.} 
\label{misidfrac_tab}
\begin{ruledtabular}
\begin{tabular}{cccccc}
            & \multicolumn{5}{c}{Tagged as}  \\
Generated   & $e^+ e^-$ & $J/\psi X$ & $J/\psi \pi^+ \pi^-$ & $J/\psi \pi^0 \pi^0$ & $J/\psi \eta$  \\
\hline
$e^+ e^-$ & - & $0.038\pm0.001$ & $<10^{-3}$ & $<10^{-3}$ & $<10^{-3}$  \\
$J/\psi \pi^+ \pi^- \to e^+ e^- \pi^+ \pi^-$ & $<10^{-3}$ & - & - & $<10^{-3}$ & $<10^{-3}$  \\
$J/\psi \pi^0 \pi^0 \to e^+ e^- \gamma \gamma \gamma \gamma$ & $<10^{-3}$ & - & $0.0015\pm0.0003$ & - & $0.0056\pm0.0005$  \\
$J/\psi \eta \to e^+ e^- \gamma \gamma$ & $<10^{-3}$ & - & $<10^{-3}$ & $<10^{-3}$ & -  \\
$J/\psi \eta \to e^+ e^- \pi^+ \pi^- \pi^0$ & $<10^{-3}$ & - & $0.104\pm0.002$ & $<10^{-3}$ & $<10^{-3}$  \\
$J/\psi \eta \to e^+ e^- \pi^0 \pi^0 \pi^0$ & $<10^{-3}$ & - & $0.0031\pm0.0004$ & $0.061\pm0.002$ & $<10^{-3}$  \\
$\chi_{c1} \gamma \to J/\psi \gamma \gamma \to e^+ e^- \gamma \gamma$ & $<10^{-3}$ & - & $<10^{-3}$ & $<10^{-3}$ & $0.0150\pm0.0008$  \\
$\chi_{c2} \gamma \to J/\psi \gamma \gamma \to e^+ e^- \gamma \gamma$ & $<10^{-3}$ & - & $<10^{-3}$ & $<10^{-3}$ & $<10^{-3}$  \\
\end{tabular}
\end{ruledtabular}
\end{table*}

The numbers of events after background subtraction are summarized in Table 
\ref{nbkgsubtr_tab}. 

\begin{table*}
\caption{Numbers of events with statistical errors and background subtraction. 
$N_{ext}$ and $N_{int}$ are the external and internal background contributions.} 
\label{nbkgsubtr_tab}
\begin{ruledtabular}
\begin{tabular}{cccccc}
   & $e^+ e^-$ & $J/\psi X$ & $J/\psi \pi^+ \pi^-$ & $J/\psi \pi^0 \pi^0$ & $J/\psi \eta$  \\
\hline
$N_{evts}$ & $5104\pm71$	& $25388\pm159$  & $4153\pm64$        & $1696\pm41$	   & $363\pm19$  \\
$N_{ext}$  & $7^{+15}_{-5}$	& $316\pm46$	 & $15^{+19}_{-10}$   & $13^{+17}_{-9}$    & $<11$ (68\% C.L.)  \\
$N_{int}$  & $-$		& $206\pm9$	 & $48\pm4$	      & $28\pm2$	   & $65\pm7$  \\
$N$	   & $5097\pm73$	& $24866\pm166$  & $4090\pm67$        & $1655\pm44$	   & $298\pm20$  \\
\end{tabular}
\end{ruledtabular}
\end{table*}


\section{Selection efficiency} \label{effic_sec}
The overall detection efficiency $\epsilon$ for each exclusive channel is 
the product of the efficiencies of the cuts used for the event selection:
\begin{equation}
\epsilon = \alpha \cdot \epsilon_{trig} \cdot \epsilon_{mee} 
              \cdot \epsilon_{EW} \cdot \epsilon_{sel} ,
\label{toteffi_fmla}
\end{equation}
where $\alpha$ is the acceptance for the \ee pair coming from charmonium 
decay, $\epsilon_{trig}$ the trigger efficiency, $\epsilon_{mee}$ the 
efficiency of the \ee invariant mass cut ($M_{ee} > 2.6$ GeV), 
$\epsilon_{EW}$ the efficiency of the EW cut ($EW_1 \cdot EW_2 > 1.5$) and 
$\epsilon_{sel}$ the final selection efficiency, which includes also the 
acceptance for all the remaining particles of the final state. 
The relevant quantity in this analysis is the ratio of the efficiencies 
for different channels. 

\subsection{Acceptance and trigger efficiency} \label{acctrig_sec}

The final state is characterized by a high invariant mass \ee pair. 
The hardware \ee trigger is based on the \v{C}erenkov counter, 
multiplicity and topology in the hodoscopes and  
multiplicity and topology and energy release in the CCAL. 

The \ee acceptance is determined by the \v{C}erenkov 
fiducial volume, which is defined as $15^\circ \leq \theta \leq 60^\circ$.

In this analysis, the product $\alpha \cdot \epsilon_{trig}$ is evaluated 
for each channel using a sample of Monte Carlo events. 

The main source of error on $\alpha \cdot \epsilon_{trig}$ is the 
uncertainty in the \ee angular distribution. 
In the reaction $\overline{p} p \to \psi^\prime \to e^+ e^-$ the \ee pair 
is distributed according to $1 + \lambda_{\psi^\prime} \cos\theta^*_e$, where 
$\theta^*_e$ is the angle between the $e^\pm$ and the beam directions 
in the center of mass system and $\lambda_{\psi^\prime}$ is the 
angular distribution parameter. 
In what follows we used $\lambda_{\psi^\prime} = 0.67 \pm 0.16$, as recently 
measured by this experiment \cite{Ambrogiani:2004uj}.

For the decay $\psi^\prime \to J/\psi \pi \pi$, the transition is 
dominated by the emission of an $L=0$ dipion \cite{Bai:1999mj}. 
In this case the \jpsi has the same polarization as the \psip and 
the angular distribution for the \jpsi decay is 
$1 + \lambda_{\psi^\prime} \cos\theta^*_e$; in the Monte Carlo 
simulated events we also assumed S wave between the dipion and the \jpsi. 

For the $J/\psi \eta$ channel, the \ee angular distribution is 
$1 + \frac{5}{4} \lambda_{\psi^\prime} - \lambda_{\psi^\prime} \cos\theta^*_e$ 
as discussed in \cite{Smith:1993}.

For the double radiative decay channels 
$\psi^\prime \to \chi_{cJ} \gamma \to J/\psi \gamma \gamma$, representing 
a background for some of the channels analyzed here, the \psip radiative 
decay angular distributions are simulated using the quadrupole amplitudes 
measured by CBAL \cite{Oreglia:1981fx}, while for the $\chi_{cJ}$ radiative 
decays, the angular distributions used in the simulation are the ones 
measured by E835 \cite{Ambrogiani:2001jw}. 

The results obtained for $\alpha \cdot \epsilon_{trig}$ are summarized 
in Table \ref{acctrigeff_tab}. 
Using these values we obtained the ratios of the 
$\alpha \cdot \epsilon_{trig}$ product between different channels, namely: 
\begin{eqnarray}
\frac{\alpha \cdot \epsilon_{trig}(J/\psi X)}{\alpha \cdot \epsilon_{trig}(e^+ e^-)} &=& 0.9579 \pm 0.0016 , \\
\frac{\alpha \cdot \epsilon_{trig}(J/\psi X)}{\alpha \cdot \epsilon_{trig}(J/\psi \pi^+ \pi^-)} &=& 1.022 \pm 0.007 , \\
\frac{\alpha \cdot \epsilon_{trig}(J/\psi X)}{\alpha \cdot \epsilon_{trig}(J/\psi \pi^0 \pi^0)} &=& 1.010 \pm 0.006 , \\
\frac{\alpha \cdot \epsilon_{trig}(J/\psi X)}{\alpha \cdot \epsilon_{trig}(J/\psi \eta)} &=& 0.83 \pm 0.03 , 
\end{eqnarray}
where the error is systematic and comes from the uncertainty in the \ee 
angular distributions. 

\begin{table}
\caption{$\alpha \cdot \epsilon_{trig}$ for different values of 
$\lambda_{\psi^\prime}$ obtained with Monte Carlo simulation. 
The $\chi_{cJ} \gamma$ channel, by radiative decay, leads to 
$J/\psi \gamma \gamma$. 
The values for the \jpsi inclusive decay are obtained as the weighted 
average of the exclusive channels.} 
\label{acctrigeff_tab}
\begin{ruledtabular}
\begin{tabular}{cccc}
Channel & $\lambda_{\psi^\prime} = 0.51$ & $\lambda_{\psi^\prime} = 0.67$ & $\lambda_{\psi^\prime} = 0.84$  \\
\hline
$e^+ e^-$            & 0.5114 & 0.4995 & 0.4893  \\
$J/\psi \pi^+ \pi^-$ & 0.4820 & 0.4681 & 0.4577  \\
$J/\psi \pi^0 \pi^0$ & 0.4831 & 0.4738 & 0.4619  \\
$J/\psi \eta$        & 0.5672 & 0.5751 & 0.5823  \\
$\chi_{c0} \gamma$   & 0.5249 & 0.5248 & 0.5233  \\
$\chi_{c1} \gamma$   & 0.5094 & 0.5068 & 0.5030  \\
$\chi_{c2} \gamma$   & 0.5026 & 0.4937 & 0.4923  \\
$J/\psi X$           & 0.4890 & 0.4785 & 0.4692  \\
\end{tabular}
\end{ruledtabular}
\end{table}

\subsection{\ee invariant mass cut efficiency} \label{meeeff_sec}
The efficiency of the \ee invariant mass cut, $m_{ee} > 2.6$ GeV, 
is $\epsilon_{mee} \simeq 96\%$, 
and affects all the channels in the same way, except $\psi^\prime \to e^+ e^-$ 
which has a higher efficiency because of the higher \ee invariant mass. 
The efficiency ratio is determined from Monte Carlo to be 
$\frac{\epsilon_{mee}(\psi^\prime \to J/\psi X \to e^+ e^- X)}{\epsilon_{mee}(\psi^\prime \to e^+ e^-)} = 0.992 \pm 0.001$. 

\subsection{Electron weight cut efficiency} \label{eweff_sec}
The presence of additional showers in the CCAL in proximity to the 
$e^\pm$ tracks modifies the CCAL shower shape for the $e^\pm$, 
resulting in a lower EW cut efficiency for these events. 
The events containing extra CCAL showers within 100 mrad of 
$e^\pm$ tracks are rejected; this allows the use of the same EW efficiency 
for all channels. 
The residual fluctuations of the $\epsilon_{EW}$ values for different 
channels, due to additional effects, have been studied with the 
Monte Carlo and are within 2\%, which has been taken as the EW cut systematic 
error. 

\subsection{Final selection efficiency} \label{finalseleff_sec}
The effect of the 100 mrad cut in the CCAL, the acceptance for exclusive reactions, 
and the kinematic fit selection efficiency for each analyzed channel 
are included in the final selection efficiency $\epsilon_{sel}$. 
This is determined from Monte Carlo event samples by 
performing the same cuts applied on data. 
The systematic error associated with the efficiency is obtained by trying 
various ${\rm prob}(\chi^2)$ cuts in the range $10^{-4}$ - $10^{-1}$ for the
$J/\psi \eta$ channel and $10^{-6}$ - $10^{-2}$ for all the other channels, 
for both data and Monte Carlo events, and observing the result fluctuations. 
The results obtained for each channel are shown in Table \ref{effisel_tab}. 

\begin{table*}
\caption{$\epsilon_{sel}$ for each decay channel. The error is systematic 
and is obtained by examining several prob($\chi^2$) cuts.} 
\label{effisel_tab}
\begin{ruledtabular}
\begin{tabular}{cccccc}
Channel & $e^+ e^-$ & $J/\psi X$ & $J/\psi \pi^+ \pi^-$ & $J/\psi \pi^0 \pi^0$ & $J/\psi \eta$  \\
\hline
$\epsilon_{sel}$ & $0.943\pm0.031$ & $0.996\pm0.011$ & $0.319\pm0.011$ & $0.229\pm0.014$ & $0.502\pm0.024$  \\
\end{tabular}
\end{ruledtabular}
\end{table*}


\section{Results} \label{results_sec}
Using the efficiency values obtained in the previous section and 
(\ref{fractevents_fmla}) we obtain: 
\begin{eqnarray}
\frac{{\cal B}(\psi^\prime \to e^+ e^-)}{{\cal B}(\psi^\prime \to J/\psi X) \cdot {\cal B}(J/\psi \to e^+ e^-)} \nonumber \\ 
= 0.206\pm0.003\pm0.008 , \\
\frac{{\cal B}(\psi^\prime \to J/\psi \pi^+ \pi^-)}{{\cal B}(\psi^\prime \to J/\psi X)} \nonumber \\ 
= 0.525\pm0.009\pm0.022 , \\
\frac{{\cal B}(\psi^\prime \to J/\psi \pi^0 \pi^0) \cdot {\cal B}(\pi^0 \to \gamma \gamma)^2}{{\cal B}(\psi^\prime \to J/\psi X)} \nonumber \\ 
= 0.292\pm0.008\pm0.019 , \\
\frac{{\cal B}(\psi^\prime \to J/\psi \eta) \cdot {\cal B}(\eta \to \gamma \gamma)}{{\cal B}(\psi^\prime \to J/\psi X)} \nonumber \\ 
= 0.0197\pm0.0013\pm0.0013 , 
\end{eqnarray}
where the first error is statistical and the second systematic. 

The number of events observed in the $J/\psi \pi^0 \pi^0$ and 
$J/\psi \eta$ decay modes could also be normalized to the events 
observed in the $J/\psi \pi^+ \pi^-$. Our result for these decay modes 
could alternatively be expressed as: 
\begin{equation}
\frac{{\cal B}(\psi^\prime \to J/\psi \pi^0 \pi^0)}
      {{\cal B}(\psi^\prime \to J/\psi \pi^+ \pi^-)} = 0.571\pm0.018\pm0.044 ,  \label{isosp_psi2pi_fmla}
\end{equation}
to be compared with the value 0.5 expected from isospin conservation, and: 
\begin{equation}
\frac{{\cal B}(\psi^\prime \to J/\psi \eta)}
      {{\cal B}(\psi^\prime \to J/\psi \pi^+ \pi^-)} = 0.095\pm0.007\pm0.007 ,  \label{psieta_psi2pi_fmla}
\end{equation}
in agreement with theoretical estimates \cite{Voloshin:1980zf,Novikov:fa}. 

From our measurement and the world averages for 
${\cal B}(\psi^\prime \to J/\psi X) = 0.557\pm0.026$, 
${\cal B}(J/\psi \to e^+ e^-) = 0.0593\pm0.0010$, 
${\cal B}(\pi^0 \to \gamma\gamma) = 0.98798\pm0.00032$ and 
${\cal B}(\eta \to \gamma\gamma) = 0.3943\pm0.0026$
\cite{Hagiwara:pw} we can derive the following values for the \psip 
branching ratios:
\begin{eqnarray}
{\cal B}(\psi^\prime \to e^+ e^-) &=& (6.8\pm0.1\pm0.4)\times10^{-3} , \\
{\cal B}(\psi^\prime \to J/\psi \pi^+ \pi^-) &=& 0.292\pm0.005\pm0.018 , \\
{\cal B}(\psi^\prime \to J/\psi \pi^0 \pi^0) &=& 0.167\pm0.005 \pm0.014 , \\
{\cal B}(\psi^\prime \to J/\psi \eta) &=& 0.028\pm0.002\pm0.002 . 
\end{eqnarray}
The results are in excellent agreement with the recent measurements 
published by BES \cite{Ablikim:2004mv} \cite{Bai:2004cg}. 
The present result is compared with the ones obtained previously by our 
experiment and E760 in Table \ref{expcompare_tab}. The larger systematic 
error in the present results comes from a more conservative evaluation of 
the kinematic fit systematic.

\begin{table*}
\caption{\psip BR measurements obtained by E835.} 
\label{expcompare_tab}
\begin{ruledtabular}
\begin{tabular}{ccccc}
    & $\displaystyle\frac{{\cal B}(\psi^\prime \to e^+ e^-)}{{\cal B}(\psi^\prime \to J/\psi X)}$ & $\displaystyle\frac{{\cal B}(\psi^\prime \to J/\psi \pi^+ \pi^-)}{{\cal B}(\psi^\prime \to J/\psi X)}$ & $\displaystyle\frac{{\cal B}(\psi^\prime \to J/\psi \pi^0 \pi^0)}{{\cal B}(\psi^\prime \to J/\psi X)}$ & $\displaystyle\frac{{\cal B}(\psi^\prime \to J/\psi \eta)}{{\cal B}(\psi^\prime \to J/\psi X)}$   \\
\hline
E760 \cite{Armstrong:pg}       & $0.0144\pm0.0008$  & $0.496\pm0.037$  & $0.323\pm0.033$  & $0.061\pm0.009$  \\
E835 \cite{Ambrogiani:2000vj}  & $0.0128\pm0.0004$  & -  & $0.328\pm0.015$  & $0.072\pm0.009$  \\
E835 (this paper)              & $0.0122\pm0.0002\pm0.0005$  & $0.525\pm0.009\pm0.022$  & $0.300\pm0.008\pm0.022$  & $0.050\pm0.006\pm0.003$  \\
\end{tabular}
\end{ruledtabular}
\end{table*}


\section{Dipion invariant mass distribution measurement} \label{dipm_sec}
\begin{figure*}
\centering
\mbox{\subfigure{\epsfig{figure=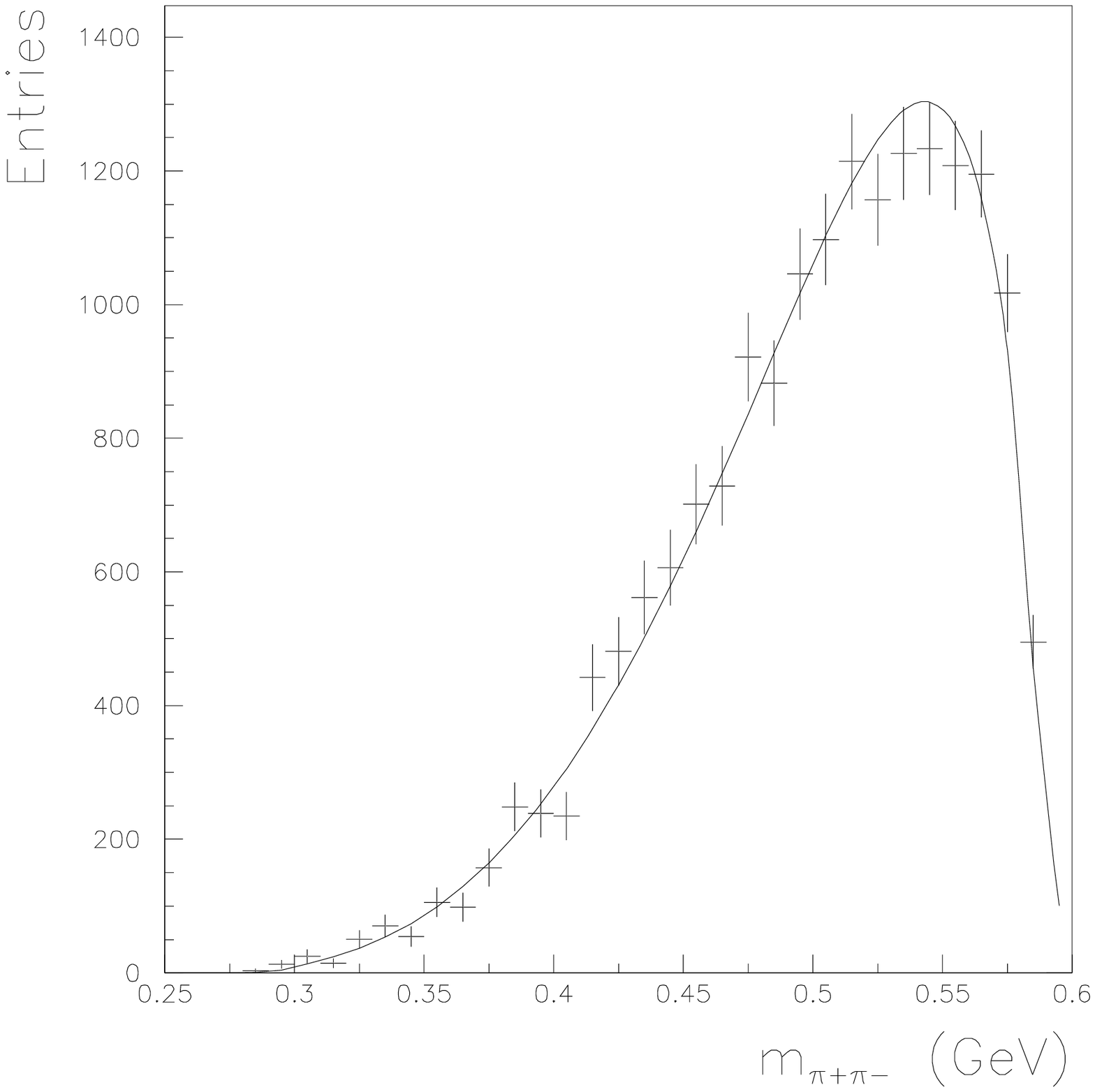,width=.45\textwidth}} \quad
      \subfigure{\epsfig{figure=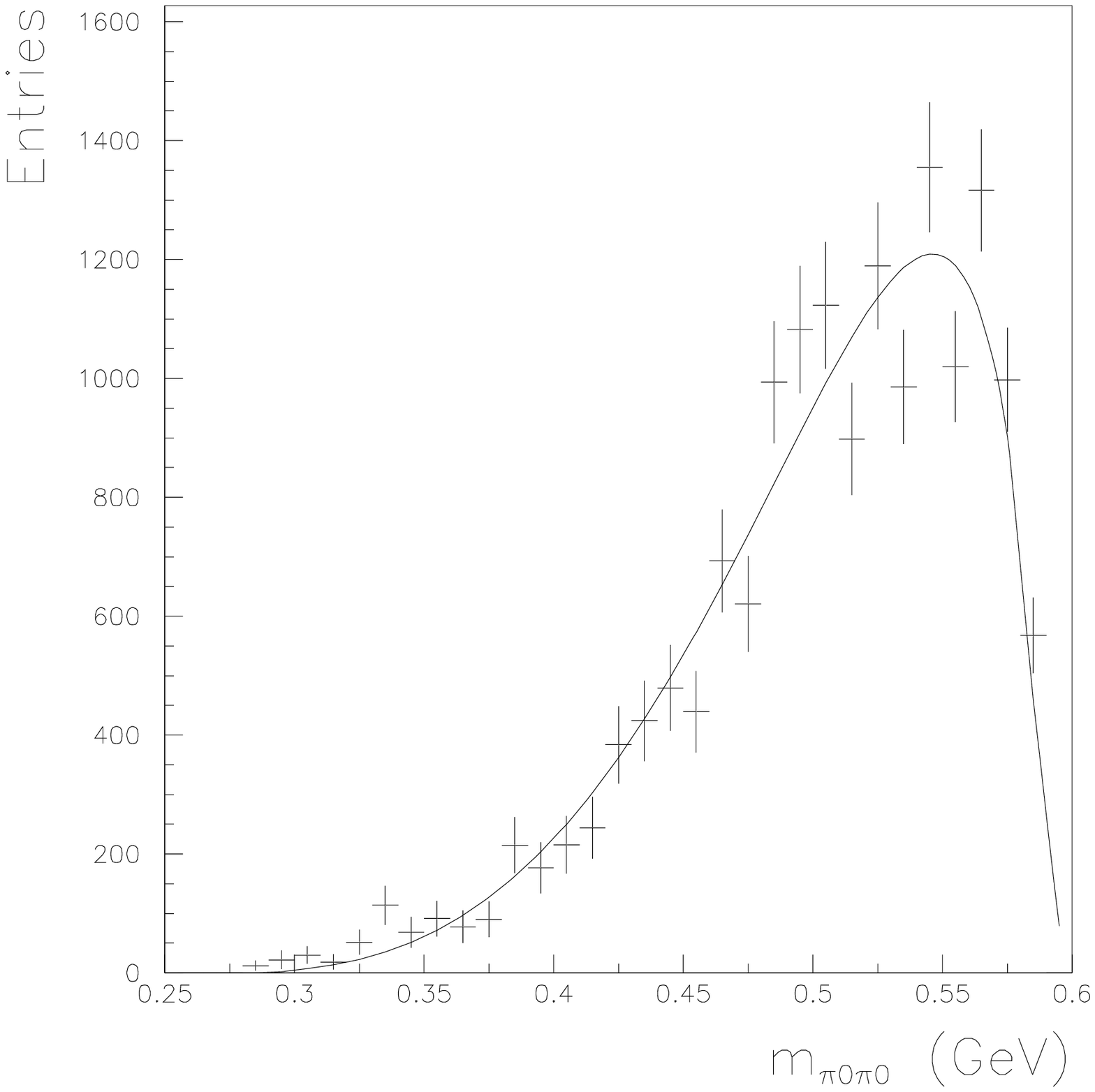,width=.45\textwidth}} }
\caption{$\pi\pi$ invariant mass distribution for (a) $J/\psi \pi^+ \pi^-$, 
(b) $J/\psi \pi^0 \pi^0$ events, corrected for the acceptance. 
The solid line is the fit with 
$P_S \cdot ( m_{\pi\pi}^2 - \Lambda_{\pi\pi} m_\pi^2 )^2$.}
\label{dipm_fig}
\end{figure*}

The dipion invariant mass ($m_{\pi\pi}$) distributions for 
$J/\psi \pi^+ \pi^-$ and $J/\psi \pi^0 \pi^0$ events, corrected for 
the detector acceptance, are shown in Figure \ref{dipm_fig}. 

A possible parametrization for the $m_{\pi\pi}$ distribution is 
\cite{Pham:1975zq}: 
\begin{equation}
\frac{d\Gamma}{dm_{\pi\pi}} \propto P_S \cdot ( m_{\pi\pi}^2 - \Lambda_{\pi\pi} m_\pi^2 )^2
\label{dipm_fmla}
\end{equation}
where the phase space $P_S$ is:
\begin{widetext}
\begin{equation}
P_S = \sqrt{ \frac{ ( m_{\pi\pi}^2 - 4 m_\pi^2)
                  [ M_{J/\psi}^4 + M_{\psi^\prime}^4 + m_{\pi\pi}^4 
                      - 2 (M_{J/\psi}^2 m_{\pi\pi}^2 + M_{\psi^\prime}^2 m_{\pi\pi}^2 + M_{J/\psi}^2 M_{\psi^\prime}^2 )] }
                {4 M_{\psi^\prime}^2} } . \nonumber
\end{equation}
\end{widetext}

The measured $m_{\pi^+\pi^-}$ and $m_{\pi^0\pi^0}$ distributions, 
corrected for the detector acceptance, are fitted to the function 
(\ref{dipm_fmla}) convoluted with their resolution as determined by 
Monte Carlo. 

The fitting procedure has been verified to reproduce the correct (input) 
result when applied to simulated events, and to perform equally well on 
the neutral and charged dipion modes. 

The results obtained for the $m_{\pi\pi}$ distribution are 
$\Lambda_{\pi^+\pi^-} = 3.31 \pm 0.15 ^{+0.35}_{-0.15}$ ($\chi^2/ndf = 32.3/29 = 1.1$) 
and $\Lambda_{\pi^0\pi^0} = 4.06 \pm 0.25 ^{+0.25}_{-0.15}$ ($\chi^2/ndf = 54.4/29 = 1.9$)
where the first error is statistical and the second systematic. 
The $m_{\pi\pi}$ distributions for $J/\psi \pi^+ \pi^-$ and 
$J/\psi \pi^0 \pi^0$ are expected to be the same.

For the evaluation of the systematic error, several fits to the data have 
been done, varying the binning, the parameters of the resolution function 
and the acceptance correction factors. The value of $\Lambda_{\pi\pi}$ 
was found to be very sensitive to the parameters of the resolution function, 
which is described by a double gaussian whose parameters are obtained 
from the Monte Carlo. The systematic errors have been obtained by varying 
the parameters of the distribution to take into account possible discrepancies 
between the data and the Monte Carlo. 
For the acceptance correction factors, some sets of values have been calculated 
under the hypothesis of $\Lambda_{\pi\pi}$ values from 2 to 6. Then the data 
have been fitted using different corrections; the $\Lambda_{\pi\pi}$ values fitted 
on data using the different correction factors show variations by a fraction 
of the statistical error, and so no systematic 
error is associated with the acceptance correction. 
We also observe that the systematic uncertainty introduced by $\Lambda_{\pi\pi}$ 
on the acceptance used in the measurement of the branching ratios of the 
$J/\psi \pi \pi$ channels is negligible with respect to the ones in Tables 
\ref{acctrigeff_tab} and \ref{effisel_tab}. 

The only other available measurement of $\Lambda_{\pi\pi}$ has been obtained by 
BES for the decay $\psi^\prime \to J/\psi \pi^+ \pi^-$, which yielded  
$\Lambda_{\pi^+\pi^-} = 4.35 \pm 0.06 \pm 0.17$ \cite{Bai:1999mj}. 
The fit to our data using this value yields a $\chi^2/ndf = 2.7$.


\begin{acknowledgments}
The authors wish to thank the staffs, engineers and technicians at 
their respective institutions for their valuable help and cooperation. 
This research was supported by the U.S. Department of Energy and by the 
Italian Istituto Nazionale di Fisica Nucleare. 
\end{acknowledgments}

\newpage 
\bibliography{psipbr_prd}

\begin{thebibliography}{9}

\bibitem{Armstrong:pg}
T.~A.~Armstrong {\it et al.}  [Fermilab E760 Collaboration],
Phys.\ Rev.\ D {\bf 55} (1997) 1153.

\bibitem{Ambrogiani:2000vj}
M.~Ambrogiani {\it et al.}  [E835 Collaboration],
Phys.\ Rev.\ D {\bf 62} (2000) 032004.

\bibitem{Goldberg:1975bz}
H.~Goldberg,
Phys.\ Rev.\ Lett.\  {\bf 35} (1975) 605.

\bibitem{Gottfried:1977gp}
K.~Gottfried,
Phys.\ Rev.\ Lett.\  {\bf 40} (1978) 598.

\bibitem{Voloshin:1980zf}
M.~B.~Voloshin and V.~I.~Zakharov,
Phys.\ Rev.\ Lett.\  {\bf 45} (1980) 688.

\bibitem{Novikov:fa}
V.~A.~Novikov and M.~A.~Shifman,
Z.\ Phys.\ C {\bf 8} (1981) 43.

\bibitem{Pham:1975zq}
T.~N.~Pham, B.~Pire and T.~N.~Truong,
Phys.\ Lett.\ B {\bf 61} (1976) 183.

\bibitem{Yan:1980uh}
T.~M.~Yan,
Phys.\ Rev.\ D {\bf 22} (1980) 1652.

\bibitem{det-paper}
G.~Garzoglio {\it et al.}  [E835 Collaboration],
Nucl.\ Instrum.\ Meth.\ A {\bf 519} (2004) 558.

\bibitem{McGinnis:gt}
D.~P.~McGinnis, G.~Stancari and S.~J.~Werkema,
Nucl.\ Instrum.\ Meth.\ A {\bf 506} (2003) 205.

\bibitem{Baldini:wg}
W.~Baldini, D.~Bettoni, R.~Calabrese, E.~Luppi, R.~Mussa and G.~Stancari,
Nucl.\ Instrum.\ Meth.\ A {\bf 449} (2000) 331.

\bibitem{geant}
GEANT - Detector Description and Simulation Tool,
http://wwwinfo.cern.ch/asd/geant/

\bibitem{Hagiwara:pw}
K.~Hagiwara {\it et al.}  [Particle Data Group Collaboration],
Phys.\ Rev.\ D {\bf 66} (2002) 010001.

\bibitem{Ambrogiani:2004uj}
M.~Ambrogiani {\it et al.}  [Fermilab E835 Collaboration],
arXiv:hep-ex/0412007.

\bibitem{Bai:1999mj}
J.~Z.~Bai {\it et al.}  [BES Collaboration],
Phys.\ Rev.\ D {\bf 62} (2000) 032002.

\bibitem{Smith:1993}
A.~J.~Smith,
Ph.~D. thesis, University of California, Irvine (1993).

\bibitem{Oreglia:1981fx}
M.~Oreglia {\it et al.},
Phys.\ Rev.\ D {\bf 25} (1982) 2259.

\bibitem{Ambrogiani:2001jw}
M.~Ambrogiani {\it et al.}  [E835 Collaboration],
Phys.\ Rev.\ D {\bf 65} (2002) 052002.

\bibitem{Ablikim:2004mv}
M.~Ablikim {\it et al.}  [BES Collaboration],
Phys.\ Rev.\ D {\bf 70} (2004) 012003.

\bibitem{Bai:2004cg}
J.~Z.~Bai {\it et al.}  [BES Collaboration],
Phys.\ Rev.\ D {\bf 70} (2004) 012006

\end{thebibliography}

\end{document}